# Electrothermal active control of preconcentrated biomolecule plugs

Sinwook Park and Gilad Yossifon*

*Faculty of Mechanical Engineering, Micro- and Nanofluidics Laboratory, Technion – Israel Institute of Technology, Technion City 3200000, Israel*

\* Corresponding author: yossifon@technion.ac.il

**Abstract**

Concentration–polarization (CP)-based biomolecule preconcentration is highly effective in enhancing the detection sensitivity, yet fails to precisely and dynamically control the location of the preconcentrated biomolecule plug to ensure overlap with the sensing region (e.g. immobilized molecular probes). Here, we used electrothermal (ET) stirring as a means of controlling the location of a preconcentrated biomolecule plug. The applied microfluidic device consisted of a Nafion membrane to induce the CP, and an array of individually addressable microscale heaters for active local ET stirring. The experimental results demonstrated that such a novel platform enabled active control of the location of the preconcentrated plug of target biomolecules, ensuring its overlap with the functionalized microparticles, ultimately yielding enhanced detection sensitivity and binding kinetics. This was demonstrated using avidin-biotin particles as a simple bead-based bioassay model.

**Keywords:** Electrothermal flow, Concentration polarization, Biomolecule preconcentration, Microfluidics

Highly efficient preconcentration of biomolecules is a critical step in bioanalysis, e.g., food testing, food/water safety monitoring, drug discovery, clinical analysis, and environmental monitoring.[1–3] Preconcentration of biomolecules enables enhancement of the detection sensitivity while minimizing sample loss, especially in samples with low-abundance biomolecules, and accelerates reaction rates and reduces detection time.[4,5] The marked advances in microfluidic technology[6,7] including strategies such as steric filtration using membrane/nanogaps[8], isotachophoresis (ITP)[9–11], dielectrophoresis (DEP)[12] and ion concentration polarization (CP) [13–21], have enabled its application toward preconcentration of biomolecules, and have improved conventional bioanalysis methods, which are both time consuming and less-sensitive.

CP is an electrokinetic phenomenon that occurs due to an imbalance of ion electromigration flux passing through an ion-permselective medium (e.g., membrane, nanochannel) upon the application of an electric field, resulting in the formation of ion-depleted and -enriched layers at opposite permselective medium interfaces.[22–26] Upon the application of electric fields in an open microchannel-membrane system, a preconcentrated plug of the charged analyte forms at the outer edge of the depletion layer due to counteracting advection and electromigrative ion fluxes. CP-based preconcentration can continuously accumulate charged molecules of a sample solution and operate under a wide range of electrolyte conditions, and is thus regarded as one of most efficient and common tools for detection of charged biomolecules in micro-scale bioanalysis.[14–17,27-34]

However, the main drawback of current CP-based preconcentration systems is the inability to precisely and dynamically control the location of the preconcentrated biomolecule plug to ensure its overlap with the surface-immobilized molecular probes (e.g., antibodies) so as to enhance detection sensitivity and binding kinetics. Currently, the location of the preconcentrated biomolecule plug is difficult to predict as it is an indirect and highly sensitive outcome of many system parameters (e.g., flow rate, voltage, channel geometry etc.), necessitating extensive precalibration to ensure an overlap with the immobilized molecular probes. Hence, there is a need for a robust technique that can control the biomolecule plug location in a direct and dynamic manner that is less sensitive to the varying operating conditions and system parameters.

One approach to achieve direct control over the length of the depletion layer, which, in turn, controls the location of the preconcentrated plug, uses embedded electrodes for local stirring of the fluid, driven by alternating-current-electro-osmosis (ACEO).[35] However, the main deficiency of the ACEO mechanism is its decreased magnitude with increasing solution conductivity.[36]

Recently, we proposed a novel means of actively controlling the diffusion layer length by inducing controlled electrothermal (ET) flow combined with natural convection.[37] ET-induced flow[38,39], which results from a combination of an externally applied electric field and temperature gradients (i.e., inducing permittivity and conductivity gradients of the electrolyte), generated by local heaters embedded at the microchannel surface, can effectively modulate the concentration-polarization layer developed within a microchannel-membrane system via local stirring of the depletion layer. Such spatio-temporal control of the depletion layer with an array of individually addressable microheaters, may also potentially control the location of a plug of preconcentrated biomolecules.

Herein, we applied active ET-driven local stirring to control both the depletion layer and the location of the preconcentrated biomolecule plug, with the specific aim of ensuring its overlap with the sensing region (e.g., surface-immobilized molecular probes), to achieve enhanced detection sensitivity and binding kinetics. As a proof of concept, we used a simplified model of avidin molecules in conjunction with biotin-coated magnetic microparticles. Although CP-based preconcentration is a well-studied technique, to the best of our knowledge, the current study is first demonstration of direct and dynamic control of the location of a preconcentrated plug.

**Active control of the pre-concentration plug by local ET stirring.** To determine the effect of ET stirring on the location of the preconcentrated plug, we fabricated an electro-thermal concentration-polarization (ET-CP) platform (Fig.1) consisting of an open microchannel-membrane system which supports net flow through the main channel, which, in conjunction with the depletion layer that is generated at the perm-selective Nafion membrane results in the formation of a preconcentrated plug. An array of individually addressable thin film microheaters was embedded upstream of the Nafion membrane.

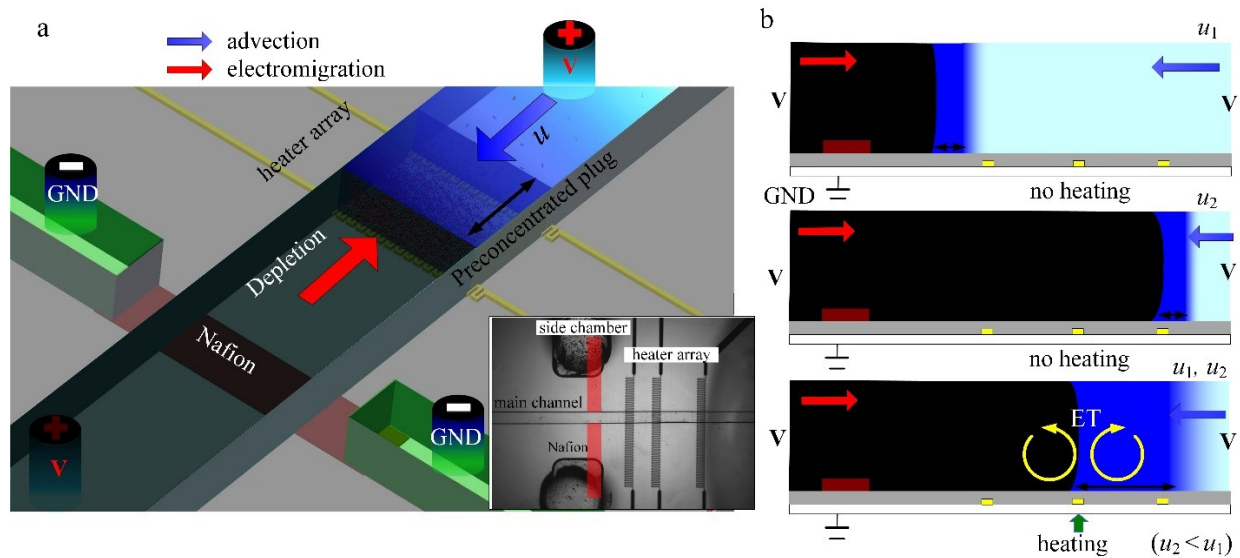

**Figure 1: Working principle of active electro-thermal (ET) flow-driven control of a concentration-polarization (CP)-based preconcentrated biomolecule plug within a microchannel-membrane device.** (a) A schematic and a microscope image of the ET-CP microfluidic platform (inset). The red arrow indicates the direction of the electromigration force acting on the negatively charged biomolecules, while the blue arrow depicts the direction of the counteracting advection both of which are necessary for the formation of the preconcentrated plug. (b) Schematic description of the dependency of the preconcentrated plug location on varying flow rates ($u_1 > u_2$) without versus with ET stirring.

Without net flow through the microchannel, the symmetric application of voltage at the inlets results in symmetric propagation of the depletion layer. Introduction of advection leads to the formation of a preconcentrated plug at the edge of the depletion layer, upstream to the Nafion interface, due to the field-gradient-focusing effect caused by counteracting advection and electromigration (see Fig.1a).[17,34] For a given microchannel geometry and ionic strength, the length of the depletion layer, which, in turn, dictates the location of the preconcentrated plug, strongly depends on the applied flow rate and the electric field intensity (see Supplementary Fig.S2, S3).

In order to precisely control the location of the preconcentrated plug, local ET stirring was applied by turning on/off local heaters that were embedded in the microchannel surface. More specifically, a step-wise heating power of 120 mW (or 150mW) was applied on the second heater in conjunction with application of an external voltage (30 V) and net flow of $u_1$ (63.8 ± 7.2 μm s$^{-1}$, Pe ~ 383) (or $u_2$ (43.2 ± 6.7 μm s$^{-1}$, Pe ~ 259)), as depicted in Fig.2a,c,e and Supplementary movie S1 (or Fig.2b,d,f and Supplementary movie S2). Under conditions of CP without ET, the equilibrium locations of the preconcentrated plug were either downstream (left) or upstream (right) to the second heater under flow rates, $u_1$ and $u_2$, respectively (Fig.2a, b, e, f). Upon activation of step-wise heating for flow rates of $u_1$ and $u_2$ at $t = 0$ s and $t = 750$ s from the moment of CP generation (i.e., application of the external DC field), respectively (Fig.2c, d), the preconcentrated plug was relocated, such that its edge sat at the center of the active heater. However, while for ET activation under $u_1$ this relocation was stable over time (Fig.2c, e), it was temporary for ET under $u_2$ (for ~200 s), with the preconcentrated plug eventually returning to its pre-ET activation equilibrium condition (Fig.2d, f). In order to retract the depletion layer that had passed the heater (pre-ET state with $u_2$; Fig.2d) and to maintain its new position for a long period of time, a higher heating power resulting in an increased ET effect, was required. These results clearly demonstrate that the location of the preconcentrated plug can be dynamically controlled by turning on/off selected ET stirring heaters. While pre-ET plug relocation from both downstream ($u_1$ case) and upstream ($u_2$ case) positions was feasible, the former allowed for better control of the length of the depletion layer.

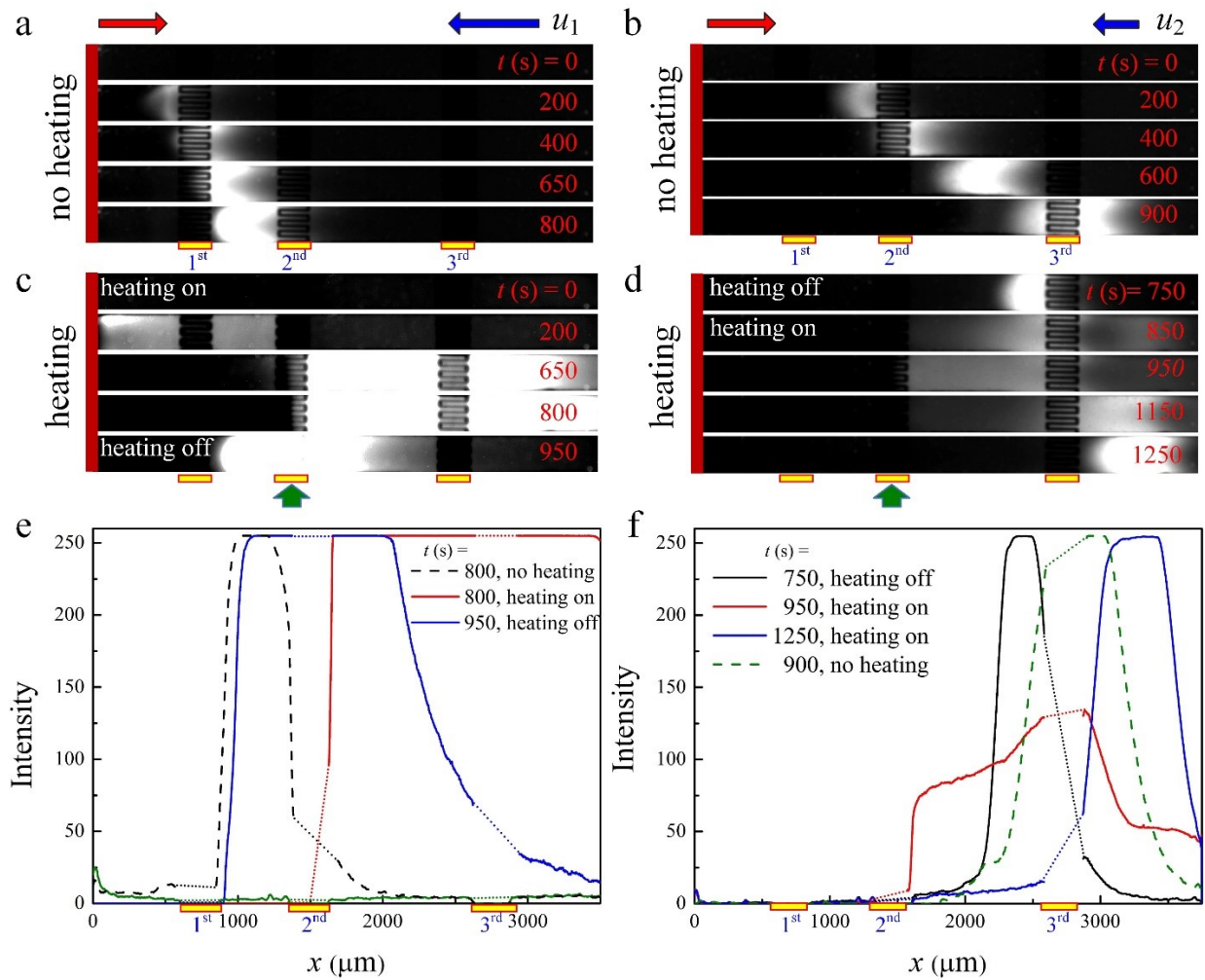

**Figure 2**: **Active control of the preconcentrated biomolecule plug location using ET-CP platform with a single-heater configuration.** (a, b) Time-transient images showing the different locations of preconcentrated plug formation under different flow rates, $u_1$ (63.8 ± 7.2 μm s$^{-1}$, Pe ~ 383) and $u_2$ (43.2 ± 6.7 μm s$^{-1}$, Pe ~ 259), without ET. (c, e) Time-transient images and their respective fluorescence intensity profiles showing the relocation of the preconcentrated plug under flow $u_1$ upon activation ($t$ = 0 s) of the second heater (120 mW) along with CP generation followed by heater deactivation at t = 800 s. (d, f) Relocation of the preconcentrated plug under flow $u_2$ upon activation of the second heater (150 mW) 750 s after CP generation. The yellow rectangles and green arrows indicate the location of all heaters and of the activated heater, respectively.

**Control of preconcentrated plug location using multiple active heaters.** To assess the effect of simultaneous activation of two heaters on the location of the preconcentrated plug, a step-wise heating power was applied via a double-heater (90 mW for each heaters) (Fig.3c, e) configuration in conjunction to a constant external voltage (30V) that generated the CP layer and forced flow (56.2 ± 3.7 µm s$^{-1}$, Pe = 337 ± 22). This was compared versus a single heater (120 mW) configuration as depicted in Fig.3b,d (Supplementary movie S3). Under CP without ET ($t_1$ = 670 s), the equilibrium location of the preconcentrated molecule plug was downstream to the second heater. While the third heater failed to relocate the preconcentrated plug, as it was too far from the equilibrium location, the second heater successfully extended the depletion region and relocated the edge of the preconcentrated plug to the center of the heater (Fig.3b). However, upon simultaneous activation of two heaters (Fig.3c), the preconcentrated plug not only relocated in between the two heaters, but also became narrower with increased intensity. Once the heaters were deactivated ($t_4$, $t_7$), the preconcentrated plug moved back to its pre-ET activation equilibrium location ($t_1$). Periodic activation of the two heaters ($t_2$, $t_6$ to $t_8$) further demonstrated the ability to dynamically control the length of the depletion layer, and thereby, the location and the width of the preconcentrated plug between the two heaters.

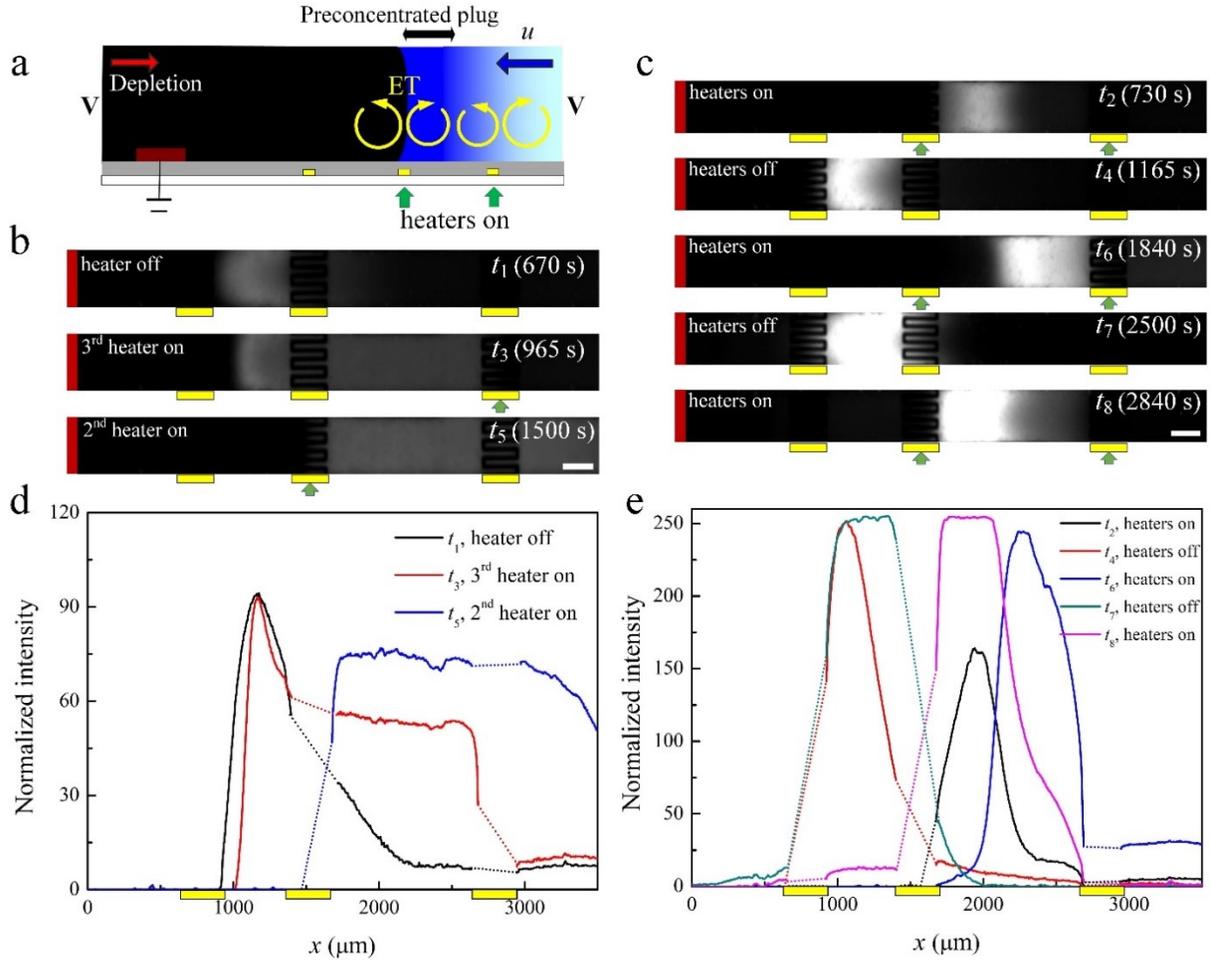

**Figure 3**: **Active control of preconcentrated plug location using a single vs. double-heater configuration.** (a) Schematic illustration of ET flow-based control of preconcentrated plug location upon activation of two heaters. The applied velocity, $u$, was 56.2 ± 3.7 μm s$^{-1}$ (Pe = 337 ± 22). Yellow, blue and red arrows indicate the direction of ET-induced vortices, advection, and electromigration, respectively. The green arrows indicate the locations of the activated heaters. (b, d) Images and corresponding intensity profiles of the location of the equilibrium preconcentrated plug formed without ET, at $t_1$ (670 s) or upon relocation of the plug by activation of either the third or second heater at $t_3$ (965 s) and $t_5$ (1500 s), respectively. (c, e) Images and corresponding fluorescence intensity profiles of the preconcentrated plug formed using a double-heater configuration at various times ($t_2$ (730 s), $t_6$ (1840 s) and $t_8$ (2840 s)) or under no-heat conditions at $t_4$ (1165 s) and $t_7$ (2500 s). Dotted lines in d, e depict an extrapolation of the intensity profile above the heaters.

The experimental results were qualitatively supported by numerical computations (Fig.4, Supplementary movie S4). The simulation results of the depletion layer and the preconcentrated plug are depicted by the co-ion ($c_2$) and third-species ($c_3$) concentration distributions, respectively. Under no-heat conditions, and a voltage of 40V and flow rate of $u$ (70 µm s$^{-1}$ or Pe = 420), the equilibrium depletion layer and preconcentrated plug were located downstream of the deactivated heaters (Fig.4). Upon activation of two heaters, the $c_2$ and $c_3$ distributions were relocatable close to the target location (between 1.33-1.8 mm – as depicted by the green rectangle in Fig.4c), which is in qualitative agreement with the experimental result in Fig.3.

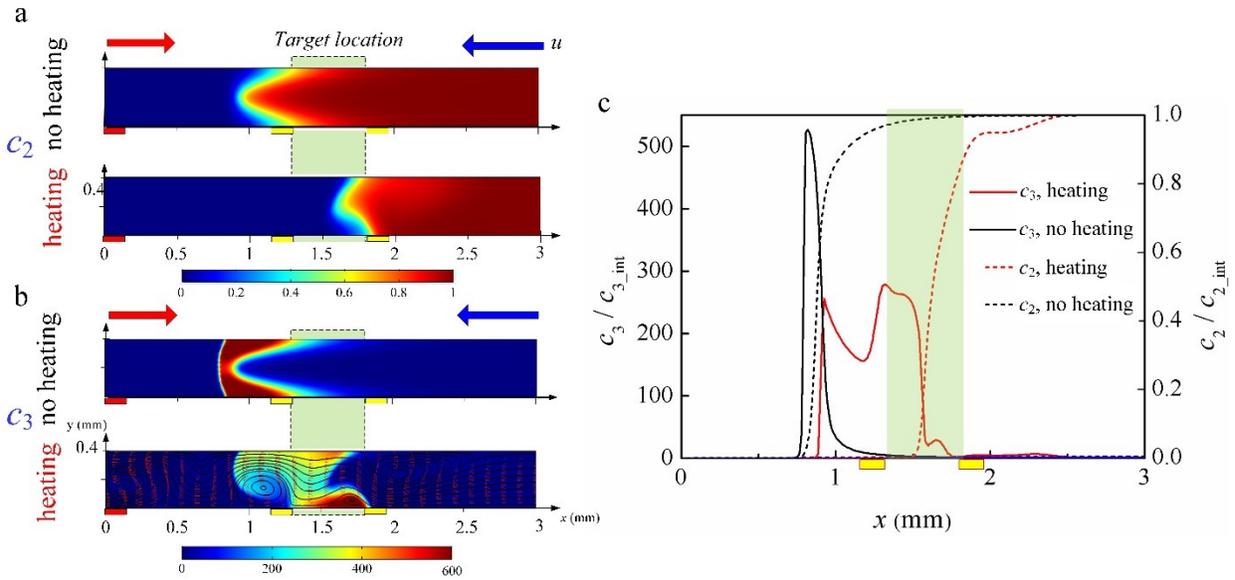

**Figure 4**: **Numerical simulation of the effect of local heating on the location of the preconcentrated plug.** (a, b) Simulation results showing equilibrium state concentration of (a) a co-ion ($c_2$) and (b) a third species ($c_3$), with/without activation (at $t$=0 s) of two heaters. The black line and red arrows indicate the flow streamlines and velocity vectors, respectively. Blue and red arrows indicate the direction of advection and electromigration force, respectively. (c) The concentration profiles of $c_2$ and $c_3$ at the center the of channel height, as obtained with/without heating.

**Binding an ET-controlled preconcentrated avidin plug to biotin-coated magnetic particles.** As a proof of concept of ET-driven enhancement of target molecule binding to immobilized molecular probes, we used the simplified model of avidin molecules in conjunction with biotin-coated magnetic microparticles. Following the magnetic trapping of biotin-coated beads (step 1 in Fig.5), a solution of 5 nM avidin was introduced under forced flow (left to right direction, Pe ~294) in conjunction with CP (30 V) and a double-heater configuration (104 mW) (see Fig.5e and Supplementary movie S5 for transient behavior of the preconcentrated avidin). Upon activation of the two heaters, the preconcentrated avidin molecule plug relocated (beyond ~600 s) from its pre-ET activation equilibrium position (t = 100 s) downstream of the first heater, to 'zone A' between the two activated heaters (Fig.5e), and stayed there for a long time, while continuously accumulating more molecules. As a result, avidin concentrations significantly increased in zone A compared to the zones B and C, reaching a significant enhancement factor of approximately two orders of magnitude (~80) relative to that achieved under the same condition without CP activation (Fig.5d, control). Consequently, avidin-biotin binding was enhanced (Fig.5d, f) with a clear correlation between the enhancement of binding and enhancement of avidin preconcentration. Taken together, these observations served as a proof of concept for enhanced binding kinetics and detection sensitivity upon combination of magnetic trapping of functionalized beads with ET-based control of the location of the preconcentrated biomolecule plug.

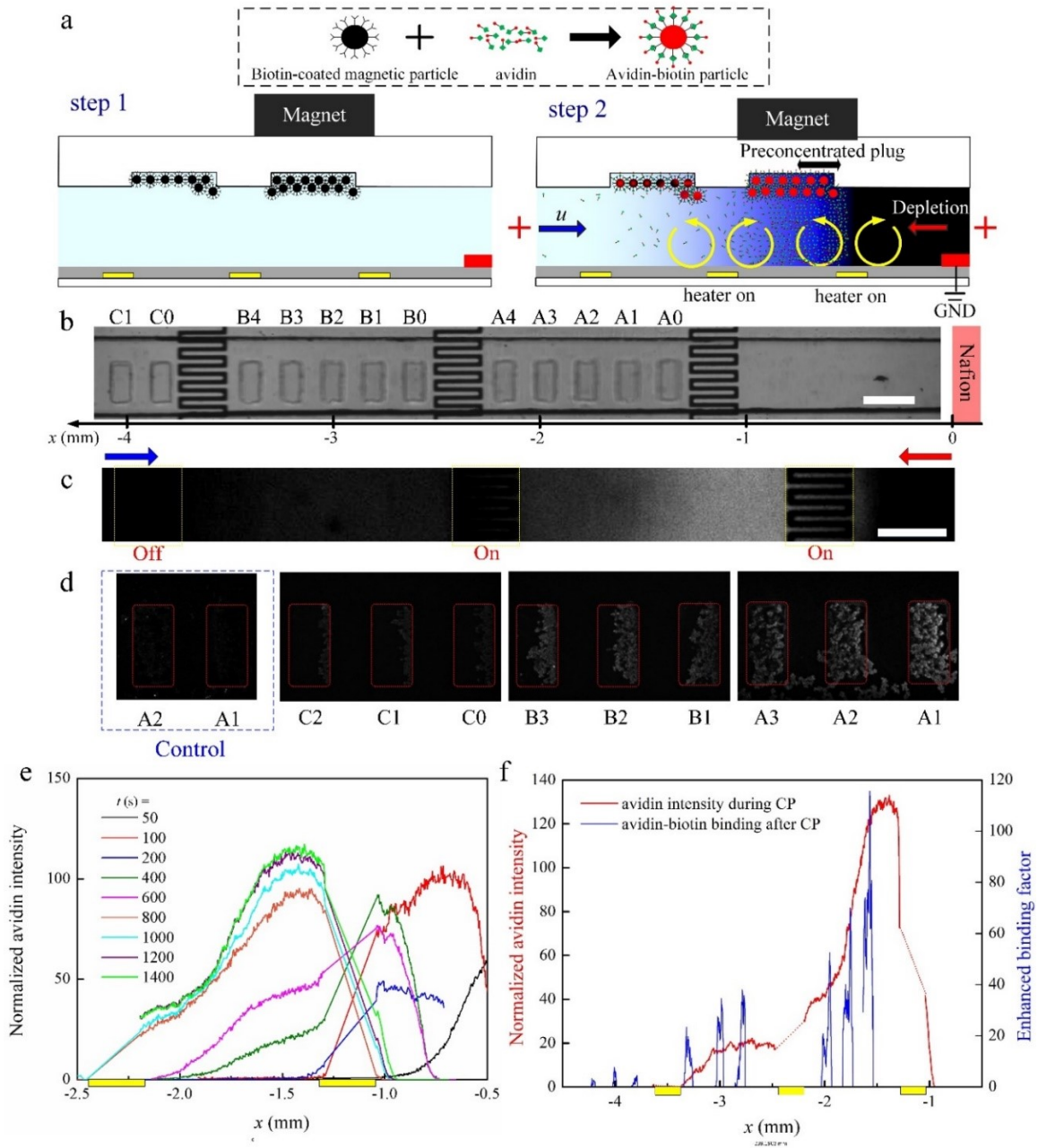

**Figure 5: ET-driven enhancement of avidin binding to biotin-coated magnetic microparticles.** (a) Schematic description of the trapping of biotin-coated magnetic particles using an external static magnet (step 1) followed by double-heater ET-controlled preconcentration of avidin molecules. Yellow, blue and red arrows indicate the direction of ET-induced vortices, forced flow, and electromigration, respectively. (b) A microscopic image of the array of heaters and trenches (intended for entrapment of the magnetic beads), embedded at the bottom and top of the microchannel, respectively. (c) Image of the fluorescent preconcentrated avidin molecules during concentration-polarization (CP) ($t \sim 1420$ s). (d) Image showing the fluorescence intensity of avidin bound to biotin-coated particles at different locations (zone A, B, C) after CP activation ($t = 1600$ s), following a washing step. The width of the red rectangles is 100μm. (e) Transient fluorescence intensity (normalized by the initial value at $t = 0$ s) distribution upon

application of CP with ET. (f) Normalized (by the initial value at $t$ =0 s) fluorescence (line-averaged) intensity distribution of the avidin molecules in (c) ($t$ =1420 s) along with the enhanced binding factor between the avidin and the biotin-coated particles taken from (d). (scale bar: 250 μm).

**Conclusions.** Here, we demonstrated, for the first time, precise and active control of the location of a preconcentrated biomolecule plug in a CP-based preconcentration device. The active control was realized by local ET stirring of the depletion layer. Without ET-driven control, the equilibrium location of the preconcentrated plug was highly dependent on the operating conditions (i.e., flow rate, DC field, etc.). In contrast, ET-driven stirring enabled controlled relocation of the preconcentrated plug via activation of a heater embedded near the target location. We also found that the simultaneous activation of two heaters further enhanced the robustness of control of bioplug location by enabling both relocation of the plug and manipulation of its width, which subsequently increased its preconcentration intensity.

The biotin-coated magnetic particles, used as a simple model of a bead-based bioassay, were immobilized using a magnetic force within an array of trenches embedded at the top of the main channel. Avidin molecules, which served as the target molecules, were then preconcentrated at desired locations by applying ET, resulting in significant enhancement (~2 orders of magnitude) of their binding to the biotin-coated particles. Compared to similar CP-based biomolecule preconcentration studies for immunoassay applications, our ET-driven active control of bioplug location, enables precise and dynamic control of the depletion layer and the associated preconcentrated plug by simply turning heaters on/off. Overlapping the preconcentrated plug with the region of immobilized antibodies is critical for enhancing binding kinetics and detection sensitivity. This novel spatio-temporal control of the depletion layer and of the preconcentrated plug, is expected to be important for many applications beyond biosensing, e.g., on-chip electrodialysis[40], where shortened diffusion lengths intensify ion transport, and CP-based desalination and pathogen rejection,[41] where control of the separation line between the brine and desalted streams is desirable.

**Methods.** *Design of an ET-CP platform with/without a trench array*. The ET-CP platform (Fig.1a) consists of a polydimethysiloxane (PDMS) main straight channel (300 μm in width, 14 mm in length and 400 μm in depth), with an embedded Nafion membrane (300 μm in width, 1mm in length) that connects the main channel to the side chambers (2 mm in diameter). An array of individually addressable heaters (20 μm in width, 240 μm in length, ~150 Ω) was embedded for generation of heat and associated temperature gradients. The electrodes were then electrically insulated from the electrolyte via a thick dielectric coating of silicon nitride (1.8 μm in thickness). Further details of the fabrication process are provided in our previous publication.[35,37]. The distances between the Nafion interface and close edges of the first, second and third heaters were 650 μm, 1400 μm and 2650 μm respectively. For the avidin-biotin binding assay, an array of rectangular trenches (100 μm in width, 12 μm in depth) was embedded on the top of the main microchannel using two-layer photolithography.[9] The developed ET-CP platforms with a trench array had a slightly different microchannel depth (350 μm) and heater locations (constant distances between heaters) as compared to the one without the trench. The array of trenches overlapped the array of heaters, resulting in three zones (A, B, C) relative to the heaters, where the distance between the Nafion interface and the close edges of the first, second and third heaters was 1045 μm, 2200 μm, 3355 μm, respectively (Fig.5b).

*Active manipulation of the preconcentrated plug in the ET-CP platform.* For CP generation, four external platinum electrodes (0.5 mm-diameter) were inserted within each of the inlets of the main channel and two side reservoirs, and connected to a power supply (Keithley 2636) with a symmetric voltage application. Details of the chip wetting and cleaning steps prior to experiments are described elsewhere[35,42]. To observe the preconcentrated plug, 5 μM of negatively charged Dylight fluorescent molecules (Dylight 488, Thermo Scientific) mixed in a ~1.2 mM KCl solution ($\sigma = 180$ μS cm$^{-1}$) served as used as a third species. Net flow was applied by a syringe pump with a withdrawal mode (KDS Legato 200 series, KD Scientific), at rates ranging from 250 nL/min and 500 nL/min. The velocities within the main channel were tracked using 2 *μm*-fluorescent particles (Thermo Scientific), at a concentration of 0.002%, resulting in Pe numbers from 185 to 387. For ET-induced flow, the second and third microheaters were activated with a constant heating power (120mW) using a DC power supply (Agilent 3612A), either separately or together. All experiments with Dylight fluorescent molecules were recorded using an Andor Neo sCMOS camera attached to a Nikon TI inverted epi-fluorescence microscope with a 4× objective lens. The measured fluorescence intensities were further analyzed by normalizing the local fluorescent dye intensity to that of initial intensity before application of the electric field.

*Numerical simulations*. The Poisson-Nernst-Planck-Stokes (PNPS) equations were solved along with the energy equation using a fully coupled, two-dimensional (2D) time-dependent model (see Supplementary for more details on the numerical simulations), using COMSOL Multiphysics 5.3. The electro-convection- and natural convection-induced flows were accounted for through the corresponding body force terms in the Stokes equation.[37,43]

*Avidin and biotin-coated magnetic particle preparation*. Commercially available 4.5 μm-diameter, biotin-coated magnetic particles (Spherotech) were diluted to a concentration of 0.025% w/v in 0.01% PBS ($\sigma = 180$ μS cm$^{-1}$). A small amount of non-ionic surfactant (0.05% v/v Tween 20 (Sigma Aldrich)) was added to minimize adhesion to the microchannel surface. Fluorescein-tagged avidin D (Vector Laboratories)[44] (5 nM in 0.01% diluted PBS) was used as the target molecule.

*Binding scheme of avidin molecules and biotin-coated magnetic particles*. To assess applicability of the system in bioassays, under the operating parameters of CP (30 V) and forced flow velocity (~ 49 μm s$^{-1}$, Pe ~ 294), Dylight fluorescent molecules diluted in 0.01% PBS were used as the preconcentrating analyte before introducing the avidin molecules. After trapping the biotin-coated magnetic particles by magnetic force exerted by external cylindrical static magnet (2mm in diameter, 5 m, 4.2 Gs, T.M.M Magnetic Technologies Ltd.), followed by a washing step of 200s, the avidin molecules were preconcentrated and incubated for 1400 sec at room temperature. During the incubation, the preconcentrated avidin plug controlled by two activated heaters, was monitored using a Nikon Eclipse Ti inverted microscope, fitted with a × 4 lens and equipped with a spinning disc confocal system (Yokogawa CSU-X1) connected to a camera (Andor iXon3). As a control, the avidin molecules were incubated in the device without preconcentration. After incubation, their binding was quantified by measuring the line-averaged fluorescence intensity along each trench. The enhancement binding factor of each trench was extracted by averaging the line intensity of the trench normalized by the averaged fluorescence intensity of the avidin bound in the trenches without CP.

# AUTHOR INFORMATION

**Corresponding author:**
* (G.Y.) E-mail: yossifon@tx.technion.ac.il.
**Author Contributions**

S.P. performed experiments, image and data analysis and numerics. G.Y. initiated research and supervised planning and execution of experiments, numerics and data analysis. All authors contributed to preparation of the manuscript.

**Notes**

The authors declare that they have filed a patent, application number 07035-P0075A, related to the ET flow driven active control of the concentration-polarization layer length in a microchannel-Nafion membrane system.

## ACKNOWLEDGMENTS

This work was supported by ISF Grant 1938/16. Fabrication of the chip was made possible through the financial and technical support of the Russell Berrie Nanotechnology Institute and the Micro-Nano Fabrication Unit.